\documentclass[10pt,aps,pra,twocolumn,showpacs,floatfix]{revtex4}
\usepackage{epsfig}
\usepackage{graphicx}
\usepackage{dcolumn}
\usepackage{longtable}
\usepackage{amsthm,amsmath}

\begin{document}

\preprint{\today}

\title{Relativistic Spectroscopy of Plasma Embedded Li-like Systems with the Screening Effects in the Two-body Debye 
Potentials}

\author{$^1$Madhulita Das, $^2$B. K. Sahoo and $^1$Sourav Pal}

\affiliation{$^1$National Chemical Laboratory, Pune - 411008, India}

\affiliation{$^2$Theoretical Physics Division, Physical Research Laboratory,
Navrangpura, Ahmedabad - 380009, India}

\begin{abstract}
The spectroscopic properties of Li atom and Li-like Ca and Ti ions in the plasma environment are investigated 
using a relativistic coupled-cluster (RCC) method. Assuming the plasma is of low density and very hot, we consider 
the Debye model with two approximations by accounting the screening effects: (i) in the nuclear potential alone and (ii) 
in both the nuclear and the electron-electron interaction potentials. First, the calculations for the energies and the 
lifetimes of the atomic states are carried out for the plasma free systems in order to check their accuracies 
following which they are investigated in the plasma environment. It is observed that screenings in the electron-electron interaction potentials 
stabilize the systems more than the case when the screenings present only in the nuclear potential. Similarly, 
the earlier predicted blue and red shifts in the $\Delta n=0$ and $\Delta n \ne 0$ transition lines (with the principal 
quantum number $n$) of the Li-like ions change significantly in the case (ii) approximation. The level-crossings 
among the energy levels are observed for the large screening effects which seem to be prominent in the states of higher orbital angular momentum. The lifetimes of many low-lying states and the relative line-intensities of 
the allowed transitions are estimated considering different plasma screening strengths.
\end{abstract}

\pacs{}

\maketitle

\section{Introduction}

The knowledge of spectral properties of the atomic systems immersed in the plasma environment is an important tool for 
plasma diagnostic processes and also have a wide range of applications in many research fields such as in the astrophysics, 
inertial confinement fusion (ICF), magnetic confinement fusion, laser-matter interaction, X-ray lasers, etc. \cite{Roger, 
Storm,Mourou, Workman, Murnane}. Usually, the spectral properties of the plasma embedded 
atomic systems change considerably in comparison to the isolated systems due to the presence of other charged particles in 
the plasma confinement. The radiations coming out from the plasma embedded atomic systems, through various excitations or 
ionization processes, play crucial roles for providing insights of the physical phenomenon happening inside the plasma. It
has been found from the laser-plasma experiments that the red shifts of the atomic emission lines are of the order of 3.7
eV \cite{Renner, Saemann}. The detailed understanding of such findings requires a proper characterization of the atomic 
spectral lines emitted from the systems. Therefore, the theoretical investigation of the atomic spectroscopy in the hot and 
dense plasma has become an exciting field of research. 
 
The plasma environment mostly consists of ions and free electrons, which introduce screening effects in the Coulomb 
potentials of the embedded atomic systems. Thus, the atomic electrons are highly influenced by the external electromagnetic 
fields compelling the atomic long range electrostatic potentials to act as short range screened potentials. In general, it 
is computationally difficult to take complete interactions into the calculations. However, the theoretical studies of the 
properties in these atomic systems are conveniently carried out by defining suitable model potentials in the atomic 
Hamiltonian by estimating the screening effects depending upon the strength of the plasma. The strength of the plasma is 
defined by a coupling parameter ($\Gamma$) that measures the interactions between the particles inside the plasma 
environment. Debye model \cite{Murillo} is the most conventional approach used for studying atomic spectroscopy in low 
density and high temperature plasma (weakly coupled plasma; i.e $\Gamma$ $<$ 1). 

 The phenomenon of reduction of ionization potential (IP) of an atom or an ion in the plasma environment is known as ionization 
potential depression (IPD) \cite{Inglis,Feynman,Ecker,Stewart}. Accurate determination of this quantity can infer many
useful information such as equation of state of plasma, radiative opacity of stellar plasma, inertial confinement 
fusion plasma, etc. Recently, experimental studies by Hoarty {\it et al.} \cite{Hoarty} and Ciricosta {\it et al.} 
\cite{Ciricosta} reveal the influence of hot and dense plasma environment in the atomic properties and the IPDs in the Al 
atom are reported. Thus, the atomic structure probes in plasma have attracted both the experimentalists and theoreticians 
to make further attempt in yielding more accurate spectroscopic data. Many calculations on H-like and He-like ions in the 
dense plasma environment have been performed \cite{Qi2003,Saha2005,Paul2009,Saha2009} for their simpler structures, but only 
a handful number of theoretical investigations are available for Li-like systems \cite{Lin2010}. Nonetheless, there have been many emission spectra observed both in the astrophysical and the laboratory plasma corresponding to high nuclear charge \cite{Boiko1980,Kelly1973,Aglitskii1974,Kononov1976,
Bromage1977,Burkhalter,Boiko1978,Fawcett,Boiko1979,Bretion,Elden}. For reliable identifications of these spectral 
lines and to comprehend better about the underlying physics of their origin, it is necessary to employ more accurate 
methods that can include both the relativistic and electron correlation effects appropriately in the theoretical 
analysis of the atomic spectral properties. 

 The primary thrust of the present work is to study theoretically the effects of the plasma screenings in the spectroscopy
of the highly ionized Ca XVIII and Ti XX ions (Li-like) and compare against the neutral Li atom. These ions are commonly 
used in the laser-plasma experiments \cite{Moreno, Clark} and are of particular interest for the astrophysical-plasma 
studies \cite{Djenize}. Before evaluating the atomic properties in the plasma environment, we determine the IPs and the 
lifetimes of the atomic states in the plasma free case to compare them with the available experimental values to validate 
our calculations. Next, we introduce the screening effects in the nuclear potential using the Debye model and then add them 
in the electron-electron interaction potential in order to find out how the results differ in these two approximations. A 
relativistic coupled-cluster (RCC) many-body theory in the Fock-space representation has been employed for the calculations. 
We also give predicted values for the ratios of the line intensities among the atomic transitions. 

  The paper is organized as follows: In Sec. \ref{sec2}, we introduce the screening models that are considered in the 
calculations of the atomic spectral properties and describe the employed RCC method briefly in Sec. \ref{sec3}. We 
give the obtained results in Sec. \ref{sec4} along with their comparison with the other available results and 
discuss them before summarizing the work in Sec. \ref{sec5}. Through out the paper, the units given for IP, excitation 
energies (EE), and fine structure (FS) splittings are in $cm^{-1}$, line strength ($S$) and the Debye screening 
length ($D$) are in atomic units (a.u.), transition rates ($A$) are in $sec^{-1}$ and lifetimes ($\tau$) of the atomic 
states are in $sec$. 

\section{Debye Model Potentials}\label{sec2}

The atoms or ions embedded in a plasma are largely perturbed by the electromagnetic fields produced by the neighboring 
ions and the freely moving electrons of the plasma. It is almost impossible to account all the possible governing 
interactions among the particles in the exact determination of the structures of the atomic systems in such 
environment. Alternatively, the effectiveness of the interactions can be approximated using appropriate models to 
be able to explain the systems within reasonable accuracies for all the practical purposes. In a hot and low density 
plasma, the atomic systems get screened due to the penetration of the slowly moving free electrons of the plasma into
the systems. Using the Poisson-Boltzmann equation, the effective potential under these conditions can be well 
approximated in the Debye model for a point nuclear system as
\begin{eqnarray}
V_{\rm{eff}}(r_i) &=& -\frac{Z}{r_{i}} e^{-r_i/D} + \sum_{j \ge i}^{N_e}\frac{1}{r_{ij}} e^{-r_{ij}/D} \nonumber \\
           &=& V_{nuc}(r_i) + \sum_{j \ge i}^{N_e} V_{ee}(r_{ij}),  
\label{veff} \\
\text{where} && D= \left [ \frac{k_{\rm{B}} T_e}{4\pi(1+Z)n_e} \right ]^{1/2}
\label{debye}
\end{eqnarray} 
is the Debye screening length, $Z$ is the nuclear charge, $N_e$ is the number of the electrons, $k_{\rm{B}}$ is the 
Boltzmann constant, $T_e$ is the plasma temperature, and $n_e$ is the plasma electron density.  

 In Eq. (\ref{veff}), $V_{nuc}(r)$ represents the screened electron-nucleus potential whereas $V_{ee}(r)$ is the 
screened electron-electron interaction potential. Due to computational difficulties, majority of the previous works 
on the spectroscopy study have considered the screenings only in $V_{nuc}(r)$ where the electron-electron potential 
$V_{ee}(r)$ is truncated at its first term of the exponential expansion for its dominant contribution 
\cite{Dray}. We refer this approximation as ``Model A'' in this work. It is also important to take into account 
the screened effects in the electron-electron interactions for large plasma strengths to achieve more realistic 
results in the search for stability of the atomic structure in the plasma medium.

Instead of considering the point nucleus in the evaluation of $V_{nuc}(r)$, we use the Fermi-charge distribution to 
take into account the finite size of the nucleus by using the expression
\begin{flushleft}
\begin{eqnarray}
 V_{nuc}(r) = -\frac{Ze^{-r/D}}{\mathcal{N}r} \times \ \ \ \ \ \ \ \ \ \ \ \ \ \ \ \ \ \ \ \ \ \ \ \ \ \ \ \ \ \ \ \ \ \ \ \ \ \ \ \  \nonumber\\
\left\{\begin{array}{rl}
\frac{1}{b}(\frac{3}{2}+\frac{a^2\pi^2}{2b^2}-\frac{r^2}{2b^2}+\frac{3a^2}{b^2}P_2^+\frac{6a^3}{b^2r}(S_3-P_3^+)) & \mbox{for $r \leq b$}\\
\frac{1}{r}(1+\frac{a62\pi^2}{b^2}-\frac{3a^2r}{b^3}P_2^-+\frac{6a^3}{b63}(S_3-P_3^-))                           & \mbox{for $r>b$}
\end{array}\right.   
\label{eq12}
\end{eqnarray}
\end{flushleft}
where the factors are 
\begin{eqnarray}
\mathcal{N} &=& 1+ \frac{a^2\pi^2}{b^2} + \frac{6a^3}{b^3}S_3  \nonumber \\
\text{with} \ \ \ \ S_k &=& \sum_{l=1}^{\infty} \frac{(-1)^{l-1}}{l^k}e^{-lb/a} \ \ \  \nonumber \\
\text{and} \ \ \ \ P_k^{\pm} &=& \sum_{l=1}^{\infty} \frac{(-1)^{l-1}}{l^k}e^{\pm l(r-b)/a} . 
\end{eqnarray}
Here, the parameter $b$ is known as the half-charge radius, $a$ is related to the skin thickness of the nucleus. These
two parameters are evaluated by
\begin{eqnarray}
a&=& 2.3/4(ln3) \nonumber \\
\text{and} \ \ \ \ \ b&=& \sqrt{\frac {5}{3} r_{rms}^2 - \frac {7}{3} a^2 \pi^2}
\end{eqnarray}
with the appropriate values of the root mean square radius, $r_{rms}$, of the nucleus.
 
 To add the screening effects in the electron-electron interactions, we adopt the approximations considered in
\cite{Hatton, Whitten, Deb, Gutierrez, Bowen} as
\begin{equation}
V_{\mathrm{eff}}(r_i)= V_{nuc}(r_i) + e ^{-r_i/D}\sum_{j \ge i}^{N_e}\frac{1}{r_{ij}}.
\end{equation}  
This approximation is labeled as ``Model B'' in the present work.
 
\section{Method of calculations}\label{sec3}

We employ the one-valence electron attachment RCC method developed by us (e.g. see Refs. \cite{bijaya1,bijaya2,bijaya3}) for
carrying out calculations of the atomic wave functions in the considered systems which is briefly described below. The 
four-component Dirac-Hamiltonian along with the effective potentials that is used for the wave function calculations of 
the plasma embedded atomic systems is given by 
\begin{equation}
H=\sum_{i=1}^{N_e} \left [ c\vec{\alpha_{i}}\cdot\vec{p}_{i}+ (\beta-1) c^{2}+V_{\mathrm{eff}}(r_{i})\right],
\label{eqDF}
\end{equation}
where $N_e$ is number of electrons of the atomic system, $\vec{\alpha}$ and $\beta$ are the Dirac matrices and $c$ is the 
velocity of light. 

 The ground states of the considered atomic systems have the $[1s^2]$ closed-shell configuration and $2s$ valence 
orbital. Also, many of the low-lying excited states in these systems have the same closed-shell configuration with an 
electron in one of the virtual orbitals instead of the $2s$ orbital. To calculate these states, we express the atomic 
wave functions in our RCC method as
\begin{eqnarray}
|\Psi_v \rangle &=& e^T \{e^{S_v}\} |\Phi_v \rangle ,
\label{eqn13}
\end{eqnarray}
where we define the reference state $|\Phi_v \rangle$ by appending the appropriate valence orbital $v$ of the 
corresponding state to the Dirac-Fock (DF) wave function of the $[1s^2]$ configuration (denoted by $|\Phi_0\rangle$). 
Here, $T$ is the core excitation operator and $\{S_v\}$ is the normal order excitation operator from core and valence to 
the virtual orbitals. Since the states have only one valence orbital in their configurations, the exponential
form $\{e^{S_v}\}$ naturally reduces to $\{1+S_v\}$ yielding the form
\begin{eqnarray}
|\Psi_v \rangle &=& e^T \{1+S_v\} |\Phi_v \rangle .
\label{eqn13a}
\end{eqnarray}

Considering the singles and doubles approximation in the RCC theory (CCSD method), the cluster operators $T$ and 
$S_v$ are defined by
\begin{eqnarray}
T &=& T_1 + T_2 \nonumber \\
\text{and} \ \ \ \ S_v &=& S_{1v} + S_{2v} .
\label{eqn14}
\end{eqnarray}

The amplitudes of these excitation operators are obtained by solving the following equations 
\begin{eqnarray}
\langle \Phi_0^K | \overline{H_N} |\Phi_0\rangle &=& \delta_{K,0} \Delta E_{corr}
\label{eqn15} \\
\text{and} \ \ \ \ \ \ \ \ \ \ \ \ \ \ \ \ \   && \nonumber \\   
\langle \Phi_{v}^K |\overline{H_N} (1+S_v)|\Phi_v\rangle &=& \langle \Phi_v^K|\delta_{K,v}+S_v|\Phi_v\rangle \Delta E_v, \ \ \ \ \ \
\label{eqn16}
\end{eqnarray}
where the values of the superscripts $K(=1,2)$ represent the (singly and doubly) excited hole-particle states, 
$\overline{H_N}=e^{-T}H_Ne^T$ is the dressed part of the normal order Hamiltonian ($H_N$). $\Delta E_{corr}$ and 
$\Delta E_v$ are the correlation energy and attachment energy (also equivalent to negative of the ionization potential
(IP)) of the valence electron $v$, respectively. 

The lifetime of a given state $f$ is defined by the reciprocal of the total transition rate of that state due to all 
possible transition channels; i.e.
\begin{eqnarray}
\tau_f &=& \frac{1} {\sum_{\text{O},i} A^{\text{O}}_{f \rightarrow i}},
\label{eqn5}
\end{eqnarray}
for the transition rate $A^{\text{O}}_{f \rightarrow i}$ due to a radiative operator $\text{O}$. The transition 
rates via various multipole channels are given by
\begin{eqnarray}
A^{\text{E1}}_{f \rightarrow i} &=& \frac {2.02613 \times 10^{18} }{\lambda^3 (2J_f+1)} S_{f \rightarrow i}^{\text{E1}}
\label{eqn1} \\
A^{\text{M1}}_{f \rightarrow i} &=& \frac {2.69735 \times 10^{13}}{\lambda^3 (2J_f+1)} S_{f \rightarrow i}^{\text{M1}}
\label{eqn2} \\
\text{and} \ \ \ \ \ \ \ \ \ \ \ \ \ \ \ \ \ \ &&  \nonumber \\
A^{\text{E2}}_{f \rightarrow i} &=& \frac {1.11995 \times 10^{18}}{\lambda^5 (2J_f+1)} S_{f \rightarrow i}^{\text{E2}} ,
\label{eqn3}
\end{eqnarray}
where $\lambda$ is the wavelength ($\AA$) and $S_{f \rightarrow i}^{\text{O}} =|\langle f || \text{O} || i \rangle|^2$ is 
the line strength due to the corresponding transition operator $\text{O}$, and $J_f$ is the total angular momentum of 
the $f$ state. E1, M1, and E2 represent the electric dipole, magnetic dipole, and electric quadrupole transition channels
respectively.

We have used Gaussian type of orbitals (GTOs) to construct the single particle orbitals. The large and small 
radial components of a Dirac orbital in this case are expressed by
\begin{eqnarray}
P_{\kappa}(r) &=& \sum_{\nu} c_{\kappa, \nu}^P r^{l_{\kappa}} e^{-\zeta_{\nu} r^2}
\label{eqn19}
\end{eqnarray}
and
\begin{eqnarray}
Q_{\kappa}(r) &=& \sum_{\nu} c_{\kappa, \nu}^Q c_{\kappa, \nu}^P r^{l_{\kappa}} \left ( \frac{d}{dr} + \frac{\kappa}{r} \right ) e^{-\zeta_{\nu} r^2},
\label{eqn20}
\end{eqnarray}
respectively, where the summation over $\nu$ stands for the total number of GTOs used in each orbital angular momentum ($l_{\kappa}$) 
symmetry with the relativistic quantum number $\kappa$, $c_{\nu}^P$ and $c_{\nu}^Q$ are the normalization constants 
and $\zeta_{\nu}$ are the guessed parameters chosen suitably for different symmetries. In order to optimize the 
exponents, we use the even tempering condition
\begin{equation}
\zeta_{\nu} = \zeta_0 \eta^{\nu-1}
\label{eqn21}
\end{equation}
with two unknown parameters $\zeta_0$ and $\eta$.

\section{Results and Discussion}\label{sec4}

 We first study the spectroscopic properties of plasma isolated atomic systems by considering $D=\infty$ in 
Eq. (\ref{eqDF}). The rationale of carrying out these calculations are to verify the accuracies of our results
by comparing them with the available experimental values. The obtained IPs and FS 
splittings of Li I, Ca XVIII and Ti XX from the CCSD method are given in Table \ref{tab1} along with the recommended
values from the national institute of science and technology (NIST) database \cite{NIST}. We found good agreements between our results with the NIST data. Due to large magnetic effects in the 
highly charged Ca XVIII and Ti XX ions, the FS splittings are large in these systems compared to the Li atom. In Tables 
\ref{tab2} to \ref{tab4}, we give the values of $S$ and $A$ from the E1, M1, and E2 channels from the excited states
to the low-lying states of Li I, Ca XVIII, and Ti XX, respectively. Using these values, we have calculated the lifetimes of 
the excited states which are given in the same table. Our theoretical estimated lifetimes of the excited states in Li I 
are in good match with the experimental values \cite{Schulze,Volz,Nagourney,Boyd}. Experimental results for the lifetimes 
of the excited states of Ca XVIII are not available. So, we compare the results with the previously reported theoretical 
results \cite{Kanti} (see Table \ref{tab3}) and find reasonably good agreement between these results as well. In fact, there 
are neither any measurements nor any theoretical values of the lifetimes of the excited states in Ti XX are known. 
Nevertheless, we foresee similar levels of accuracies in these results, given in Table \ref{tab4}, based on the calculations 
in Li I and Ca XVIII.

\begin{figure*}
\includegraphics{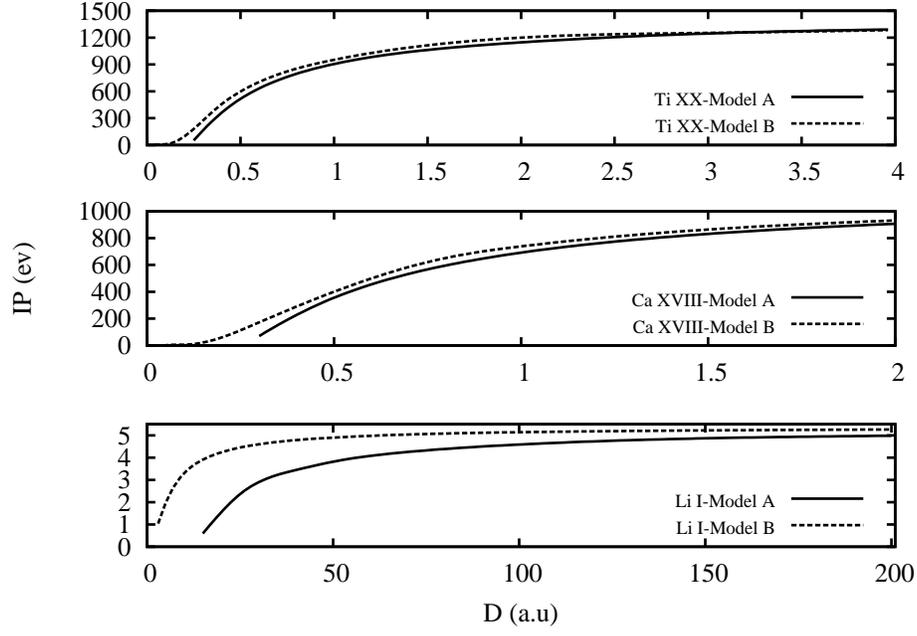}
\caption{Variation of ionization potentials with Debye screening length ($D$).}
\label{fig1}
\end{figure*}

\begin{figure*}
   \includegraphics{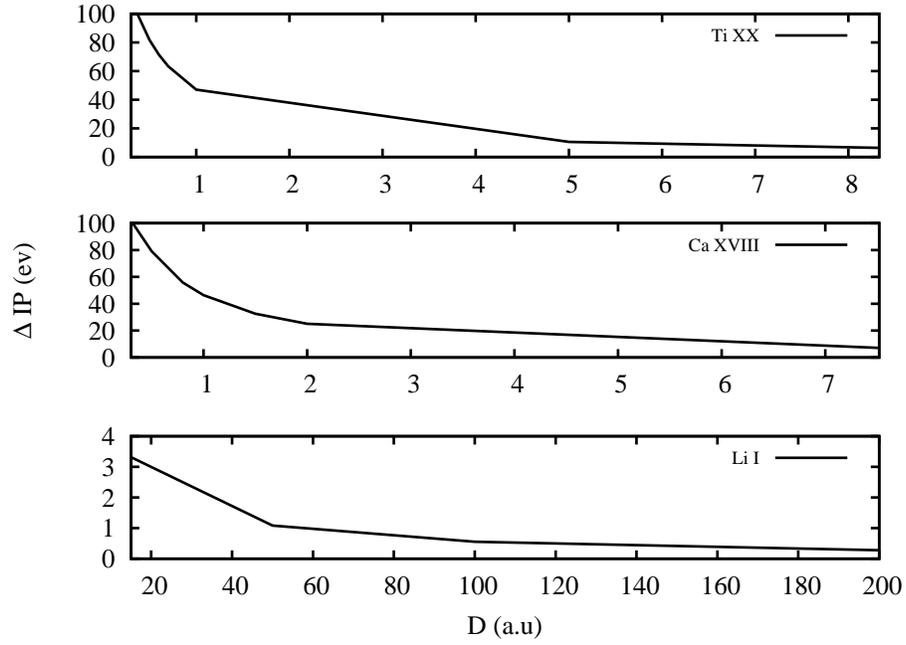}
\caption{Differences in the ionization potentials from Model B and Model A with Debye screening length ($D$).}
\label{fig2}
\end{figure*}

  After establishing the accuracies of various quantities of our interest in the isolated atomic systems, we now 
proceed in presenting these properties of the plasma embedded systems by considering finite values of $D$ in 
Eq. (\ref{eqDF}). Unlike the isolated system, the IP of the plasma embedded system is no longer stable and it 
changes depending upon the strength of the plasma environment. As can be seen from Eq. (\ref{debye}) that $D$ is a 
function of the plasma electron density ($n_e$) and plasma temperature ($T_e$). It is, therefore, possible to generate 
various plasma conditions by varying these parameters. For this purpose, we have varied the values of $D$ from 0.5 to 
200 a.u., 0.045 to 7.52 a.u., and 0.04 to 8.33 a.u. in the calculations of the above quantities in Li I, Ca XVIII and 
Ti XX, respectively. In Fig \ref{fig1}, we plot variation of the IPs of the ground states of Li I, Ca XVIII and Ti XX 
against the $D$ values considering both Model A and Model B approximations. As expected, the IPs decrease with 
decrease in $D$ values in both the models which is one of the unique characteristics of the plasma embedded atomic systems as 
referred as IPD or continuum lowering \cite{Inglis,Feynman,Ecker,Stewart}. In Model A, the lower values of $D$ around which 
the systems exist are 14 a.u., 0.3 a.u., and 0.25 a.u. in Li I, Ca XVIII and Ti XX, respectively, and below these critical
values of $D$, the systems are supposed to be unbounded. After introducing the screenings in the repulsive interactions
through Model B, the above critical values change from 14 a.u. to 0.4 a.u., 0.3 a.u. to 0.043 a.u., 0.25 a.u. to 0.04 
a.u., respectively, in Li I, Ca XVIII and Ti XX. It is evident from Fig. \ref{fig1} that the IPs obtained in both 
the models are almost same for large $D$ values but they differ for smaller values of $D$ (large plasma couplings). To
show it more prominently, we plot the differences of IPs from Model B and Model A ($\Delta \mbox{IP} = \mbox{IP}_{\mbox{B}}
- \mbox{IP}_{\mbox{A}}$) in Fig. \ref{fig2} against the $D$ values. It is obvious from this figure that $\Delta$IP
shows slow variation for large $D$ values while there are drastic differences and are positive for the smaller values of 
$D$ (large screening). This indicates high stability of the systems in the presence of screening effects through
the electron-electron repulsion. 

\begin{figure*}
   \includegraphics{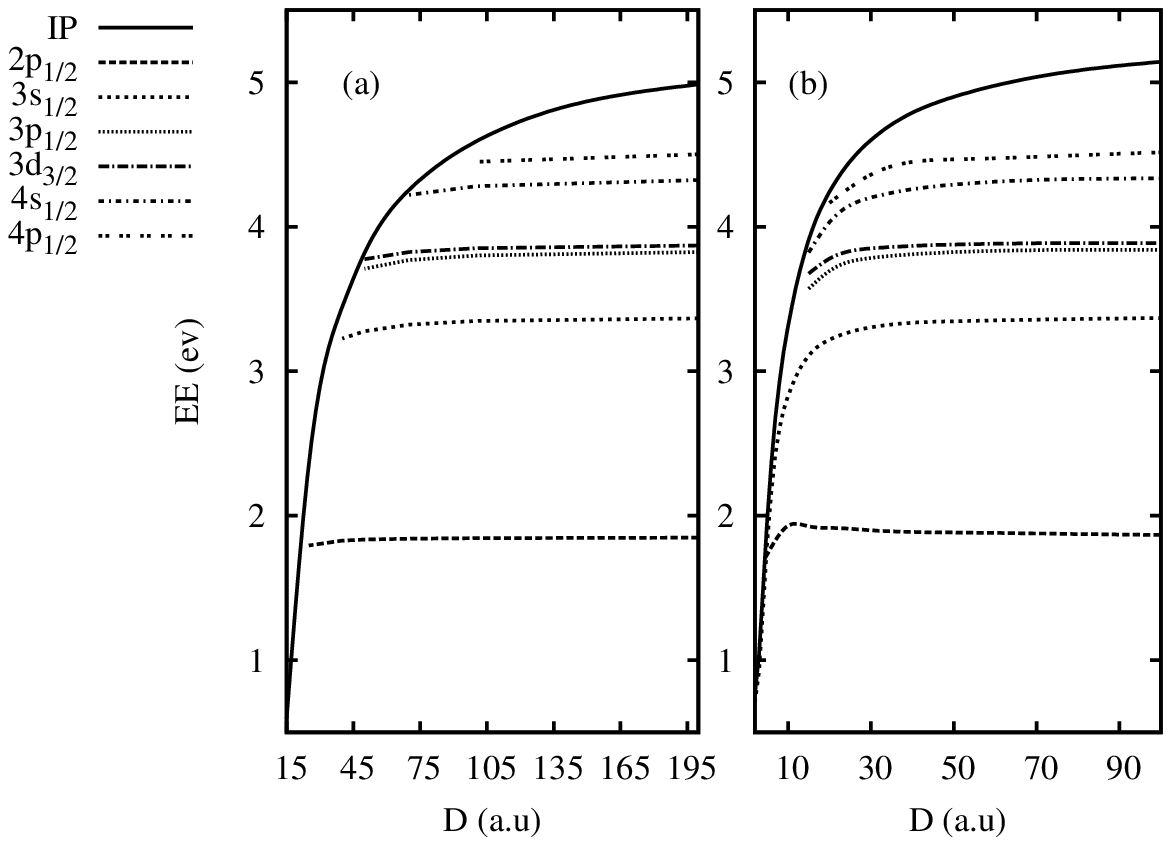}
\caption{Variation in the excitation energies of Li I with Debye screening length obtained from Model A shown in panel (a) 
and from Model B shown in panel (b).}
\label{fig3}
\end{figure*}

\begin{figure*}
\includegraphics{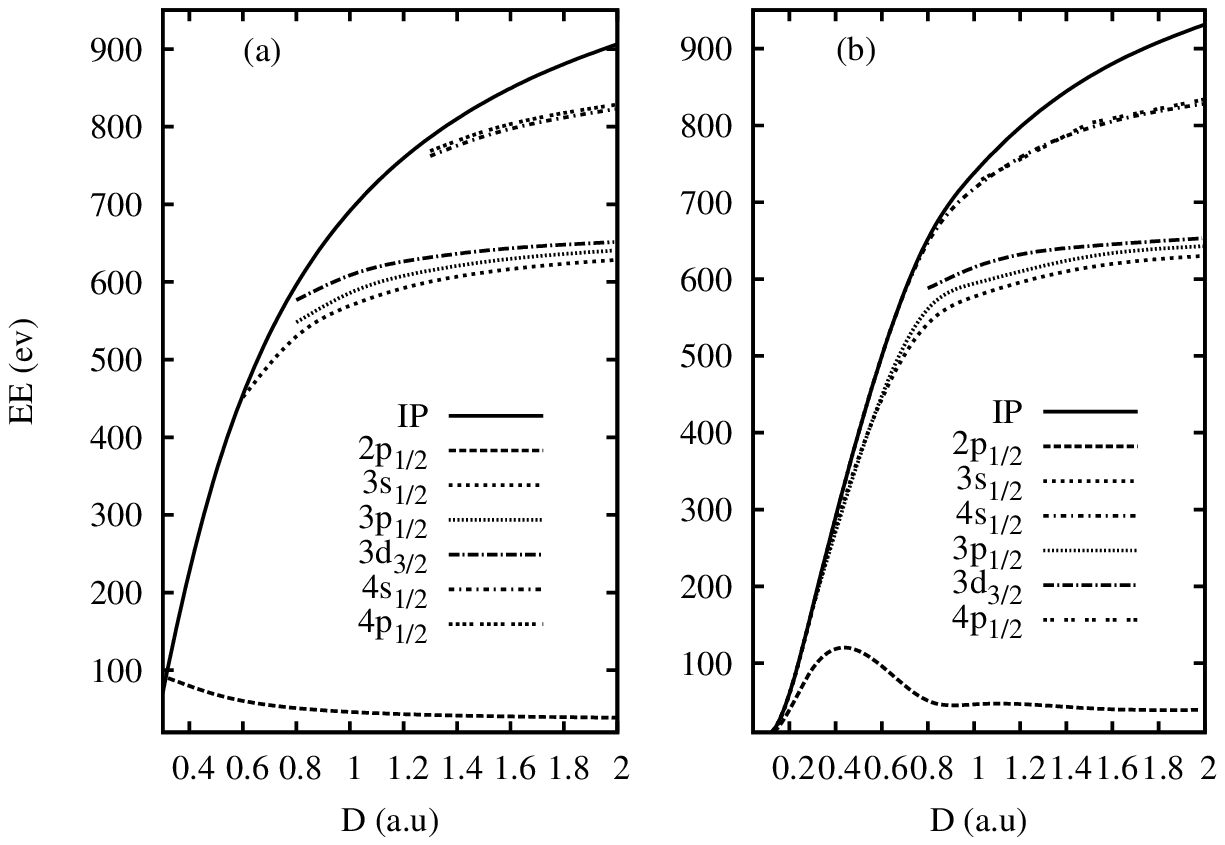}
\caption{Variation in the excitation energies of Ca XVIII with Debye screening length obtained from Model A shown in panel (a) 
and from Model B shown in panel (b).}
\label{fig4}
\end{figure*}

\begin{figure*}
\includegraphics{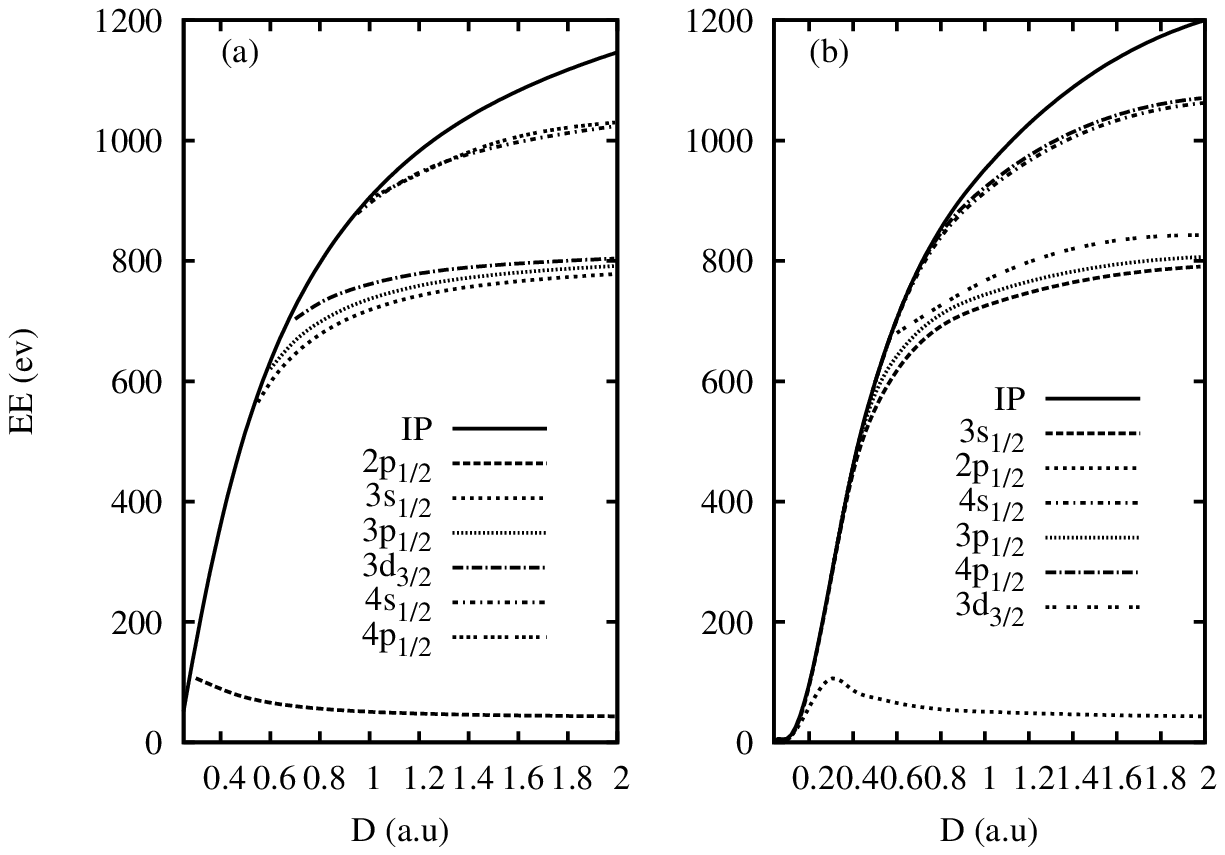}
\caption{Variation in the excitation energies of Ti XX with Debye screening length obtained from Model A shown in panel (a) 
and from Model B shown in panel (b).}
\label{fig5}
\end{figure*}

We also plot the variations in EEs of the first six low-lying excited states of Li I, Ca XVIII and Ti XX against the
$D$ values in Figs. \ref{fig3}, \ref{fig4}, \ref{fig5}, respectively, for both Model A and Model B. First, we discuss
the effects of plasma on EEs in the above systems using Model A. Fig. \ref{fig3} shows that the EEs of the 
2p$_{1/2,3/2}$, 3p$_{1/2,3/2}$, 3d$_{3/2,5/2}$, 4s$_{1/2}$, and 4p$_{1/2}$ states in Li atom  decrease with decrease 
in $D$ values and finally, the states are merged into the continuum. Similar features are also observed in the highly charged 
Ca XVIII and Ti XX ions except for the $2p$ states (see Figs. \ref{fig4} and \ref{fig5}). In these states, the EEs of 
the $2p$ states increase towards the strong screening strengths. Such signatures were also reported earlier in the H-like 
systems \cite{Qi}, He-like systems \cite{Ray}, and Li-like systems \cite{Mdas}. Therefore, Model A suggests that there are 
blue shifts in the 2s-2p transitions in the plasma embedded Ca XVIII and Ti XX ions while all the EEs in Li I show red 
shifts for the reducing values of $D$. The figures also indicate that the embedded systems lose their capability to hold the 
bound states with increasing in plasma strengths and hence, possess only a finite number of energy levels as compared to 
their counter isolated ones.

\begin{figure*}
\includegraphics{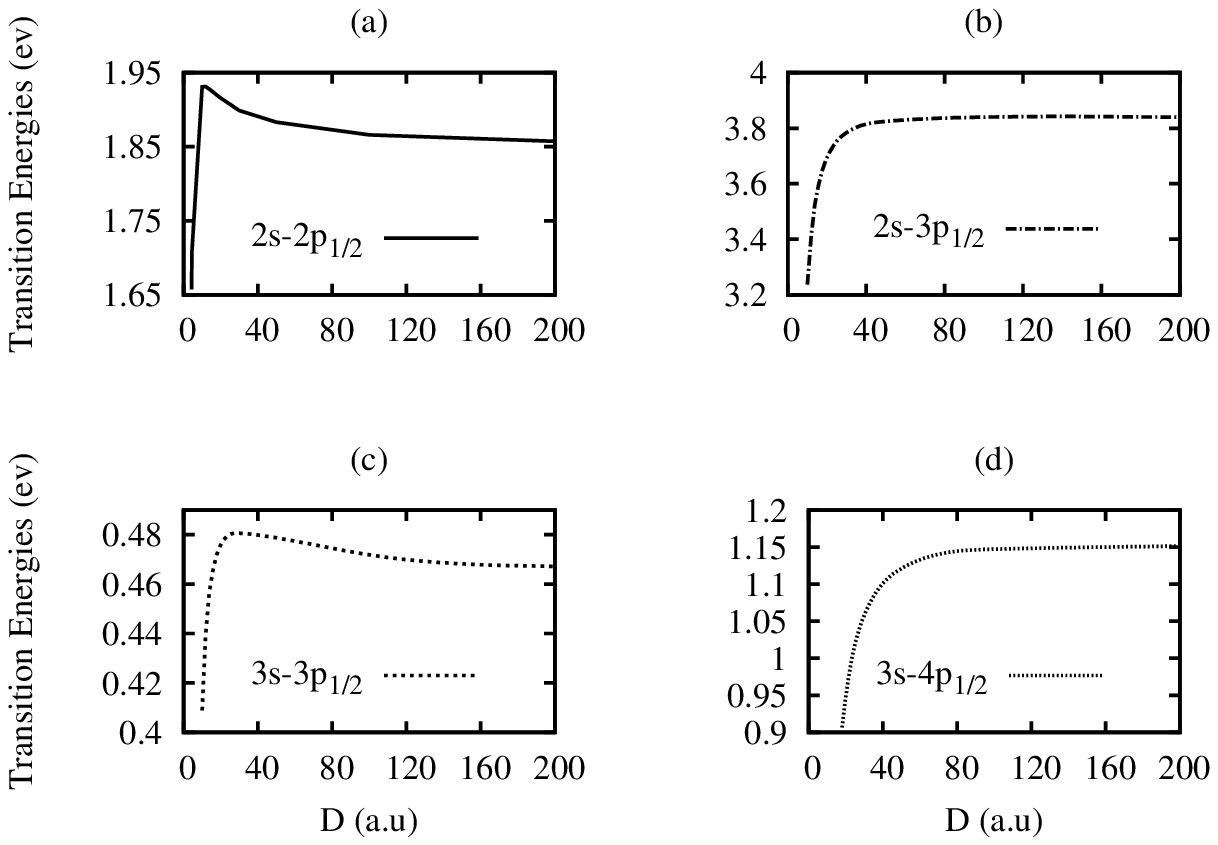}
\caption{Variation in the transition energies with Debye screening length in Li I using Model B. Panels (a) and (b) 
correspond to the 2s$_{1/2}$-2p$_{1/2}$ and 2s$_{1/2}$-3p$_{1/2}$ transitions, respectively, and panels (c) and (d) 
correspond to the 3s$_{1/2}$-3p$_{1/2}$ and 3s$_{1/2}$-4p$_{1/2}$ transitions, respectively.}
\label{fig6}
\end{figure*}

Now coming to the EEs obtained from Model B, it is clear from Figs. \ref{fig3}, \ref{fig4} and \ref{fig5} that the 
EEs of all the states except for the $2p$ states decrease with decreasing in $D$ values. Thus, these transitions show 
red-shifts and finally, move to continuum at the critical values of $D$. The EEs of the $2p$ states show different behavior; they are blue-shifted for larger the 
$D$ values (smaller screening), and then gradually become red-shifted towards the smaller $D$ region (higher screening). 
To realize these behaviors extensively, we plot the EEs of the $2s-2p_{1/2}$, $2s-3p_{1/2}$, $3s-3p_{1/2}$, and 
$3s-4p_{1/2}$ transitions of Li I and Ca XVIII against the $D$ values in Figs. \ref{fig6} and \ref{fig7}, respectively,
by considering Model B. It is clear from these figures that the EEs of the $2s-3p$ and $3s-4p$ transitions decrease 
with decrease in $D$ values as per the findings in Model A \cite{Mdas}. However, the quantities for the $2s-2p$ and $3s-3p$ transitions increase initially with decreasing in $D$ values and then fall suddenly in the lower $D$ region. Similar trends are also observed in the Ti XX ion. We infer from this behavior that the $\Delta n \neq 0$ transition spectra are red-shifted with decreasing in $D$ values for both Model A and Model B, however the 
$\Delta n = 0$ transition spectra are red-shifted in the Li atom and blue-shifted in the Ca XVIII and Ti XX ions with 
decreasing $D$ values for Model A. In contrast, the spectra for $\Delta n = 0$ transitions show blue-shifts at the 
higher values of $D$ and red-shifts for the lower values of $D$ in all the atomic systems embedded in plasma when 
Model B is taken into account. This indicates that the effects of the two-electron screenings become less important
for higher values of $D$ (lower screening strengths) for which the results of Model A and Model B show nearly same behavior,
but these screening effects play significant roles to influence the results at the lower values of $D$ (higher screening 
strength) by producing a hump like structure. It is worth pointing out here that similar features have also been observed
by Bowen {\it et al}. in the He-like systems \cite{Bowen}. 

  Figs. \ref{fig3}, \ref{fig4} and \ref{fig5} also disclose that there appear to be changes in the energy level 
structures in the high screening regions with Model B with respect to the isolated systems. In Li I, it shows energy 
level crossings between the $2p$ and $3s$ states near to the continuum edge. The $2p$ states move to continuum towards
very lower $D$ values whereas the $3s$ state still exists there. Similar level crossings are also seen in the Li-like 
Ca and Ti ions near to the high screening regions. In the ionized systems, the level crossings become more obvious. To 
demonstrate these level-crossing processes in the plasma embedded systems, we give the Grotrian energy levels diagram,
particularly in Ca XVIII, for few selective values of $D$ in Fig. \ref{fig8}. This figure shows the change in the 
sequence of energy levels with the decreasing in values of $D$. For lower screening lengths, there found to be 
sequence alternative of the energy levels. As seen in this figure, the lower $3d_{3/2,5/2}$ levels of Ca XVIII ion
moved to the continuum around $D=0.3$ a.u. even when the $4s_{1/2}$ state is still bounded with the ion. This implies that
there are level crossings between the $4s$ and $3d$ states. There are also cross-overs among the $4s$, $3s$, $2p$, and
$3p$ states in this ion such as its energy level sequence becomes as $2s$, $3p_{1/2,3/2}$, $2p_{1/2,3/2}$, $3s$, and 
$4s$. A similar trend is also noticed in the Li-like Ti ion. Fig. \ref{fig8} also illustrates that the states with 
$n=3$ in Ca XVIII ion are nearly degenerate for the higher values of $D$, but the differences among the $3s$, $3p$ and $3d$ 
states increase with the decreasing values of $D$. Gradually towards the high screening lengths, the cross-overs occur 
among the $3s$, $3p$ and $3d$ energy levels, probably, owing to the weaker attraction of bound electrons by the nucleus. It is, however, found from Fig. \ref{fig8} that the FS splittings between the states are 
least affected by the plasma screening. Indeed, we find that the states with larger orbital angular momentum are more 
affected by the screenings leading to the remarkable change in the energy level structures of the plasma embedded atomic 
systems near the continuum edge.

 Since the energy spectra of the considered atomic systems are affected by the plasma environment, it is obvious that 
the radiative properties such as lifetimes of the excited states in these systems are different than their corresponding
isolated systems. By calculating the line strengths for various allowed and forbidden transitions, we evaluate the 
lifetimes of the excited states in the considered systems with different values of $D$. The variations in the 
lifetimes of the $2p_{1/2}$, $3s_{1/2}$, $3p_{1/2}$, $3d_{3/2}$ and $4s_{1/2}$ states in Li I and Ca XVIII for 
different $D$ values are listed in Table \ref{tab5}. We plot them against the $D$ values in Fig. \ref{fig9} from
which we found that the lifetimes of the $2p$ states in Li I decrease slowly with increasing plasma strengths. 
Similar trends are also seen for the $3p$ states in the Na atom by an earlier study \cite{Chaudhuri}. Moreover, the lifetimes of the $3s$, $3d$, and $4s$ states show sharp rise in the high screening region.

  The line intensity ratio of multiplet components in an atomic system is an important tool to determine various 
plasma parameters which are useful for both the laboratory and astrophysical plasma studies 
\cite{Keenan,Porquet,Brage,Ellis}. Under the local thermodynamic equilibrium (LTE), the line intensity ratio of plasma embedded atomic systems in the optically thin plasma is given by \cite{Konjevic}
\begin{equation}
\frac{I_1}{I_2} = \left( \frac{\lambda_{nm}}{\lambda_{ki}} \right ) \left ( \frac{g_kA_{ki}}{g_nA_{nm}} \right ) e^{-(E_k-E_n)/k_BT_e},
\end{equation}
where $I_1$, $I_2$ are the line intensities and $\lambda_{ki}$, $\lambda_{nm}$ are the wavelengths of the corresponding 
$k \rightarrow i$ and $n \rightarrow m$ transitions, respectively, of the atomic system. Here $gA$s are the weighted 
transition probabilities and $E_k$, $E_n$ are the energies of the respective upper $k$ and $n$ states, respectively.

 The variation in the line intensity ratios between the $(2p ^{2}P_{1/2}-3d ^{2}D_{3/2})$ and $(2p ^{2}P_{3/2}-3s 
^{2}S_{1/2})$ allowed transitions of Ca XVIII and Ti XX ions with respect to $D$ for Model B are plotted in Fig. \ref{fig10}. The outer 
plot in Fig. \ref{fig10} is for the Ca XVIII ion and the inset plot is given for the Ti XX ion. These calculations 
are carried out at $T_e$ of 600 eV and 800 eV for Ca XVIII and Ti XX ions, respectively. It is evident from this figure that 
the variation in the intensity ratios of the above transitions increase rapidly with the $D$ values. Similar features were 
also reported earlier in the H-like \cite{Qi} and Be-like \cite{Saha2006} systems. Our computed line intensities ratio 
between the above transitions with $n_e=2 \times 10^{21}$ and $T_e=600$ eV for Ca XVIII is coming out to be 9.42 against the 
observed value 7.8 \cite{Moreno}. Similarly, we obtain this value as 9.60 for Ti XX with $n_e=1 \times 10^{21}$ and $T_e =  
800$ eV compared to the experimental result 8.4 \cite{Moreno}. Although these results agree qualitatively, but the 
discrepancies between our estimated values from the observations can be the consequences of the approximations used
in the Debye model for which we suggest consideration of more accurate theoretical calculations. One of the suitable 
approaches to minimize the differences in these results would be to include the exact form of the screening term in the 
two-body interaction potentials.

\section{Concluding remarks}\label{sec5}

  We have investigated the effects of plasma confinement on the electronic structure and spectroscopic properties of 
the Li atom and highly charged Li-like Ca XVIII and Ti XX ions with two different approximations in the Debye model using 
a relativistic couped-cluster method. The important findings from this work are summarized pointwise as follows:

\begin{enumerate}

\item In the presence of plasma confinement, the ionization potentials of the embedded systems decrease with decreasing in 
Debye lengths. Finally, at particular Debye length, the energy levels of the embedded systems merge into the continuum.

\item It is found that inclusion of the screening effects in the electron-nucleus interaction potentials destabilize the 
systems. In contrast, inclusion of the screenings in the electron-electron potentials counteract these effects by lowering 
the energies of the embedded systems, hence providing more stability to the systems. 

\item With Model A approximation, all the transition spectra in the Li atom show red shifts with increasing in plasma 
strengths. But in case of Ca XVIII and Ti XX ions, the spectra corresponding to the $ns-np$ transitions show blue shifts 
and the $ns-n'p$ transitions exhibit red-shifts with increasing in plasma strengths. However, the systems display some 
peculiar behavior in the presence of screenings in the electron-electron interactions through Model B approach. It is found 
that the $\Delta$n $\neq$ 0 transition spectra of the plasma embedded systems are red shifted with increasing in plasma 
strengths and the feature is unchanged for both Model A and Model B. In case of Model B, the $\Delta$n $=$ 0 transitions 
show blue shifts at the higher screening lengths and red shifts for lower screening lengths due to the presence of the 
screenings in the electron-electron interactions. 

\item In the presence of screening effects in both the electron-nucleus and the electron-electron interactions, the sequence 
of energy levels change for lower Debye lengths in comparison to the plasma isolated systems. The level-crossings between the states become more prominent in the highly ionized systems.  

\item For higher screening lengths, the states with the same principal quantum number remain degenerate but the differences 
in the energy levels are widen for decreasing values of $D$. 

\item Due to the strong influence of the screenings in the energy spectra of the embedded atomic systems, the lifetimes of 
the excited states change in comparison to their counter plasma free systems.

\end{enumerate}

Our reported spectral properties will be useful in the studies of both the laboratory and astrophysical plasma for 
identifying the spectral lines in the considered atomic systems and will motivate the experimentalists to verify
the theoretical findings.

\section*{Acknowledgments}
This work has been supported by Council of Scientific and Industrial Research, India. A part of the computations were 
carried out using 3TFlop HPC cluster at Physical Research Laboratory, Ahmedabad.

 \begin{table*}[t]
\caption{Comparison of our calculated ionization potentials (IPs), excitation energies (EEs) and fine-stricture splittings (FS) in Li I, Ca XVIII and Ti XX  with
the national institute of science and technology (NIST) data \cite{NIST} in $cm^{-1}$.}\label{tab1}
\begin{ruledtabular}
\begin{center}
\begin{tabular}{c c c c c c c c c}
  &\multicolumn{2}{c}{Li} &  &\multicolumn{2}{c}{Ca XVIII} &  &\multicolumn{2}{c}{Ti XX } \\
\cline{2-3}\cline{5-6}\cline{8-9}   & Present         &  NIST   &  &    Present &  NIST  &  & Present    &  NIST      \\          
\hline
 \multicolumn{9}{c}{IP}   \\
  & 43482.67  &  43487.11  &  &   9340603.58  &  9337690  &  & 11499573.80 & 11495470 \\

 \multicolumn{9}{c}{EE}   \\

2p$_{1/2}$ & 14909.23   &  14903.66               &  &   290372.20      &     290057            &  & 323898.05   &  323521  \\
2p$_{3/2}$ & 14909.19   &  14904                  &  &   333108.78      &     330918     &  & 388627.13   &  385666  \\

3s$_{1/2}$ & 27202.15   &  27206.12               &  &   5259328.13     &     5276800     &  & 6468820.72  & 6465990 \\
3p$_{1/2}$ & 30923.83   &  30925.38               &  &   5339743.89     &     5338500     &  & 6558617.61  & 6555620 \\
3p$_{3/2}$ & 30923.81   &  30925.38               &  &   5352361.91     &     5350200                  &  & 6577736.86  & 6574040 \\

3d$_{3/2}$ & 31277.94        & 31283.08                &  &   5382956.95     &     5381200                  &  & 6612097.87  & 6608120 \\
3d$_{5/2}$ & 31277.95        & 31283.12                &  &   5386772.95     &     5384200                  &  & 6617915.63  & 6613920 \\

4s$_{1/2}$ & 35007.66        & 35012.06                &  &   7063599.12     &     7059274.31               &  & 8691461.10  & 8687020\\
4p$_{1/2}$ & 36466.32        & 36469.55                &  &   7096670.50     &     7112700                  &  & 8728421.68  & 8724670 \\
4p$_{3/2}$ & 36466.31        & 36469.55                &  &   7101981.07     &     7118100                  &  & 8736469.11  & 8732440\\

 \multicolumn{9}{c}{FS}   \\        
 2p$_{1/2} - $2p$_{3/2}$        & 0.04        &  0.34                   &  &   42736.58       &     40861     &  & 64729.08    &  62145   \\
 3p$_{1/2} - $3p$_{3/2}$       & 0.02          &  0.0                    &  &   12618.02       &     11700                    &  & 19119.25    & 18420 \\
 3d$_{3/2} - $3d$_{5/2}$    & 0.01            & 0.04                    &  &   3816.00        &     3000                     &  & 5817.76     & 5800 \\
 4p$_{1/2} - $4p$_{3/2}$     & 0.01            & 0.0                     &  &   5310.57        &     5400                     &  & 8047.43     & 7770\\
\end{tabular}
\end{center}
\end{ruledtabular}
\end{table*}
  
\begin{table}
\caption{Line strengths ($S$), transition probabilities ($A$) and lifetimes ($\tau$ in $ns$) of the excited states in Li atom. 
Numbers given in the square brackets represent powers of 10.}
\begin{ruledtabular}
\begin{tabular}{lcccc}
Transition & $S_{k \rightarrow i}$ & $A_{k \rightarrow i}$& \multicolumn{2}{c}{$\tau_k$}  \\
\cline{4-5}\\
 $k \xrightarrow{\rm{O}}i$ &                            &                  & This work & Experiment \\
\hline
 & & \\
$2p_{1/2} \xrightarrow{E1}  2s_{1/2}$  & 11.019    &  3.70[7]       &  27.031  & 27.102 \cite{Schulze} \\

$2p_{3/2} \xrightarrow{E1}  2s_{1/2}$  & 22.024    &  3.70[7]       &  27.047  & 27.102 \cite{Schulze} \\
   $\ \ \ \ \ \ \ \xrightarrow{M1}  2p_{1/2}$  & 1.33     &  6.05[-16]      &                                 \\
%   $\ \ \ \ \ \ \ \xrightarrow{E2}  2p_{1/2}$  & ?     &  ?      &                                 \\
$3s_{1/2}$  $\xrightarrow{M1}  2s_{1/2}$  & 8.1[-9] &  2.21[-6]       &  29.82  & 29.72 \cite{Volz}     \\
   $\ \ \ \ \ \ \ \ \xrightarrow{E1}  2p_{1/2}$ & 5.936    &  1.12[7]      &    \\
   $\ \ \ \ \ \ \ \ \xrightarrow{E1}  2p_{3/2}$ & 11.877    &  2.24[7]       &                                \\
$3p_{1/2} \xrightarrow{E1}  2s_{1/2}$  & 0.033     &  9.97[5]       &  210.82  & 203 \cite{Nagourney} \\
%   $\ \ \ \ \ \ \ \ \xrightarrow{M1}  2p_{1/2}$ & ?  & ?        &                                  \\          
   $\ \ \ \ \ \ \ \ \xrightarrow{M1}  2p_{3/2}$ & 3.6[-8]     &  1.99[-6]       &                                  \\          
   $\ \ \ \ \ \ \ \ \xrightarrow{E2}  2p_{3/2}$ & 448.385   &  2.64[1]       &                                  \\ 
   $\ \ \ \ \ \ \ \ \xrightarrow{E1}  3s_{1/2}$ & 71.746    &  3.75[6]       &                                    \\ 
$3p_{3/2} \xrightarrow{E1}  2s_{1/2}$  & 0.067     &  1.0[6]       &  210.62  &                       \\
   $\ \ \ \ \ \ \ \ \xrightarrow{M1}  2p_{1/2}$ & 3.2[-8]     & 8.80[-7]       &                                  \\
   $\ \ \ \ \ \ \ \ \xrightarrow{E2}  2p_{1/2}$ & 448.410   &  1.32[1]       &                                 \\
   $\ \ \ \ \ \ \ \ \xrightarrow{M1}  2p_{3/2}$ & 4.0[-10]     &  1.19[-8]       &                                  \\          
   $\ \ \ \ \ \ \ \ \xrightarrow{E2}  2p_{3/2}$ & 448.246   &  1.32[1]       &                                   \\ 
   $\ \ \ \ \ \ \ \ \xrightarrow{E1}  3s_{1/2}$ & 143.484   &  3.75[6]       &                                  \\ 
%   $\ \ \ \ \ \ \ \ \xrightarrow{M1}  3p_{1/2}$ &  ?  &   ?      &                                  \\ 
%   $\ \ \ \ \ \ \ \ \xrightarrow{E2}  3p_{1/2}$ &  ?  &   ?      &                                  \\ 
$3d_{3/2} \xrightarrow{E2}  2s_{1/2}$ & 301.130  & 2.52[2]       &  14.58  & 14.60 \cite{Schulze}    \\
%   $\ \ \ \ \ \ \ \ \xrightarrow{E2}  2s_{1/2}$ & ?     &  ?      &                                 \\
   $\ \ \ \ \ \ \ \ \xrightarrow{E1}  2p_{1/2}$ & 25.721     &  5.71[7]      &                                 \\
   $\ \ \ \ \ \ \ \ \xrightarrow{E1}  2p_{3/2}$ & 5.145      & 1.14[7]       &                                 \\
   $\ \ \ \ \ \ \ \ \xrightarrow{E2}  3s_{1/2}$ & 11.3[3]  &  3.56[-1]        &                                   \\
%   $\ \ \ \ \ \ \ \ \xrightarrow{M1}  3s_{1/2}$ & ?  &  ?        &                                   \\
   $\ \ \ \ \ \ \ \ \xrightarrow{E1}  3p_{1/2}$ & 136.954    & 3.08[3]       &                                  \\
   $\ \ \ \ \ \ \ \ \xrightarrow{E1}  3p_{3/2}$ & 27.389     &  6.16[2]      &                                  \\                     
$3d_{5/2} \xrightarrow{E2}  2s_{1/2}$  & 451.743    &  2.52[2]       &  14.58  &                       \\
   $\ \ \ \ \ \ \ \ \xrightarrow{E1}  2p_{3/2}$ & 46.293     & 6.86[7]       &                               \\
   $\ \ \ \ \ \ \ \ \xrightarrow{E2}  3s_{1/2}$ & 16.9[3]  & 3.56[-1]       &                                \\
   $\ \ \ \ \ \ \ \ \xrightarrow{E1}  3p_{3/2}$ & 246.510    & 3.70[-1]       &                                 \\
   $\ \ \ \ \ \ \ \ \xrightarrow{M1}  3d_{3/2}$ & 2.400      &  2.04[-18]      &                                 \\          
%   $\ \ \ \ \ \ \ \ \xrightarrow{E2}  3d_{3/2}$ & ?      &  ?      &                                 \\          
$4s_{1/2} \xrightarrow{M1}  2s_{1/2}$  & 2.8[-8]      &  1.62[-4]      &  56.06  & 56 \cite{Boyd}        \\
   $\ \ \ \ \ \ \ \ \xrightarrow{E1}  2p_{1/2}$ & 0.420      &  3.46[6]      &                                  \\
   $\ \ \ \ \ \ \ \ \xrightarrow{E1}  2p_{3/2}$ & 0.841      & 6.92[6]       &        &                         \\
   $\ \ \ \ \ \ \ \ \xrightarrow{E1}  3p_{1/2}$ & 36.042     & 2.49[6]       &    \\
   $\ \ \ \ \ \ \ \ \xrightarrow{E1}  3p_{3/2}$ & 72.086     &  4.97[7]      &    \\        
   $\ \ \ \ \ \ \ \ \xrightarrow{E2}  3d_{3/2}$ & 8.5[3]   &  3.45[-1]      &    \\         
%   $\ \ \ \ \ \ \ \ \xrightarrow{M1}  3d_{3/2}$ & ?   &  ?      &    \\         
   $\ \ \ \ \ \ \ \ \xrightarrow{E2}  3d_{5/2}$ & 12.8[3]  &  5.17[-1]      &    \\
\end{tabular}
\end{ruledtabular}
\label{tab2}
\end{table}

\begin{table}
\caption{Line strengths ($S$), transition probabilities ($A$) and lifetimes ($\tau$ in $ps$) of the excited states in Ca XVIII 
ion. Numbers given in the square brackets represent powers of 10.}
\begin{ruledtabular}
\begin{tabular}{lcccc}
Transition & $S_{k \rightarrow i}$ & $A_{k \rightarrow i}$& \multicolumn{2}{c}{$\tau_k$}  \\
\cline{4-5}\\
 $k \xrightarrow{\rm{O}}i$ &                            &                  & This work & Others \cite{Kanti} \\
\hline
 & & \\
$2p_{1/2} \xrightarrow{E1}  2s_{1/2}$  & 5.28[-2]    &  1.31[9] & 763 & 753 \\

$2p_{3/2} \xrightarrow{E1}  2s_{1/2}$  & 1.07[-1]    &  2.00[9] & 501 & 504 \\
   $\ \ \ \ \ \ \ \xrightarrow{M1}  2p_{1/2}$  & 1.330     &  7.00[2]   &                                 \\
   $\ \ \ \ \ \ \ \xrightarrow{E2}  2p_{1/2}$  & 6.35[-3]  &  2.54[-2]   &                                 \\
$3s_{1/2}$  $\xrightarrow{M1}  2s_{1/2}$  & 1.04[-5] &  2.04[4]       & 1.08 & 1.09    \\
   $\ \ \ \ \ \ \ \ \xrightarrow{E1}  2p_{1/2}$ & 2.42[-3]  &  3.01[11]  &    \\
   $\ \ \ \ \ \ \ \ \xrightarrow{E1}  2p_{3/2}$ & 5.13[-3]  &  6.22[11]  &                                \\
$3p_{1/2} \xrightarrow{E1}  2s_{1/2}$  & 1.52[-2]  &  2.35[12]  &  0.4258  & 0.4286 \\
   $\ \ \ \ \ \ \ \ \xrightarrow{M1}  2p_{1/2}$ & 8.39[-7]  & 1.46[3]        &                                  \\          
   $\ \ \ \ \ \ \ \ \xrightarrow{M1}  2p_{3/2}$ &  3.99[-5] & 6.76[4]        &                                  \\          
   $\ \ \ \ \ \ \ \ \xrightarrow{E2}  2p_{3/2}$ &  4.78[-3]  &  8.42[8]     &                                  \\ 
   $\ \ \ \ \ \ \ \ \xrightarrow{E1}  3s_{1/2}$ &   3.23[-1] &  1.70[8]   &                                    \\ 
$3p_{3/2} \xrightarrow{E1}  2s_{1/2}$  & 2.96[-2]  &  2.30[12]  &  43.4  & 43.7                      \\
   $\ \ \ \ \ \ \ \ \xrightarrow{M1}  2p_{1/2}$ & 1.39[-5]   & 1.22[4]       &                                  \\
   $\ \ \ \ \ \ \ \ \xrightarrow{E2}  2p_{3/2}$ & 4.56[-3]   &  4.24[8]       &                                 \\
   $\ \ \ \ \ \ \ \ \xrightarrow{M1}  2p_{3/2}$ & 3.25[-5]     &  2.77[4]       &                                  \\          
   $\ \ \ \ \ \ \ \ \xrightarrow{E2}  2p_{3/2}$ & 4.71[-3]   &  4.20[8]       &                                   \\ 
   $\ \ \ \ \ \ \ \ \xrightarrow{E1}  3s_{1/2}$ & 6.49[-1]   &  2.65[8]       &                                  \\ 
   $\ \ \ \ \ \ \ \ \xrightarrow{M1}  3p_{1/2}$ &  1.332  &   1.80[1]    &                                  \\ 
   $\ \ \ \ \ \ \ \ \xrightarrow{E2}  3p_{1/2}$ &  2.35[-1]  &   2.11[-3]  &                                  \\ 
$3d_{3/2} \xrightarrow{E2}  2s_{1/2}$ & 1.52[-6]  & 1.60[3]  &  0.1432  & 0.1434 \\
   $\ \ \ \ \ \ \ \ \xrightarrow{E2}  2s_{1/2}$ & 1.74[-2]     &  2.20[9]      &                                 \\
   $\ \ \ \ \ \ \ \ \xrightarrow{E1}  2p_{1/2}$ & 8.71[-2]     &  5.83[12]      &                                 \\
   $\ \ \ \ \ \ \ \ \xrightarrow{E1}  2p_{3/2}$ & 1.77[-2]      & 1.16[12]       &                                 \\
   $\ \ \ \ \ \ \ \ \xrightarrow{E2}  3s_{1/2}$ & 1.51[-1]  &  1.22[2]        &                                   \\
   $\ \ \ \ \ \ \ \ \xrightarrow{M1}  3s_{1/2}$ & 4.01[-8] &  5.11[-4]        &                                   \\
   $\ \ \ \ \ \ \ \ \xrightarrow{E1}  3p_{1/2}$ & 4.18[-1]    & 1.71[7]       &                                  \\
   $\ \ \ \ \ \ \ \ \xrightarrow{E1}  3p_{3/2}$ & 8.33[-2]  &  1.21[6]      &                                  \\                     
$3d_{5/2} \xrightarrow{E2}  2s_{1/2}$  & 2.61[-2]    &  2.21[9] &  0.1432  & 0.1444                      \\
   $\ \ \ \ \ \ \ \ \xrightarrow{E1}  2p_{3/2}$ & 1.59[-1]     & 6.93[12]       &                               \\
   $\ \ \ \ \ \ \ \ \xrightarrow{E2}  3s_{1/2}$ & 2.28[-1]  & 1.43[2]       &                                \\
   $\ \ \ \ \ \ \ \ \xrightarrow{E1}  3p_{3/2}$ & 7.52[-1]    & 1.03[7]       &                                 \\
   $\ \ \ \ \ \ \ \ \xrightarrow{M1}  3d_{3/2}$ & 2.398      &  5.99[-1]      &                                 \\          
   $\ \ \ \ \ \ \ \ \xrightarrow{E2}  3d_{3/2}$ & 5.14[-2]      &  7.76[-7]      &                                 \\          
$4s_{1/2} \xrightarrow{M1}  2s_{1/2}$  & 3.62[-6]      &  1.72[4]  &  1.607  & 1.612        \\
   $\ \ \ \ \ \ \ \ \xrightarrow{E1}  2p_{1/2}$ & 3.81[-4]      &  1.20[11]      &                                  \\
   $\ \ \ \ \ \ \ \ \xrightarrow{E1}  2p_{3/2}$ & 7.98[-4]      & 2.47[11]       &        &                         \\
   $\ \ \ \ \ \ \ \ \xrightarrow{E1}  3p_{1/2}$ & 1.61[-2]     & 8.36[10]       &    \\
   $\ \ \ \ \ \ \ \ \xrightarrow{E1}  3p_{3/2}$ & 3.39[-2]     &  1.72[11]      &    \\        
   $\ \ \ \ \ \ \ \ \xrightarrow{E2}  3d_{3/2}$ & 2.10[-2]   &  1.57[7]      &    \\         
%   $\ \ \ \ \ \ \ \ \xrightarrow{M1}  3d_{3/2}$ & ?   &  ?      &    \\         
   $\ \ \ \ \ \ \ \ \xrightarrow{E2}  3d_{5/2}$ & 3.18[-2]  &  2.36[7]      &    \\
\end{tabular}
\end{ruledtabular}
\label{tab3}
\end{table}

\begin{table}
\caption{Line strengths ($S$), transition probabilities ($A$) and lifetimes ($\tau$ in $ps$) of the excited states in Ti XX 
ion. Numbers given in the square brackets represent powers of 10.}
\begin{ruledtabular}
\begin{tabular}{lccc}
Transition & $S_{k \rightarrow i}$ & $A_{k \rightarrow i}$& $\tau_k$  \\
 $k \xrightarrow{\rm{O}}i$ &              &           & This work \\
\hline
 & & \\
$2p_{1/2} \xrightarrow{E1}  2s_{1/2}$  & 4.15[-2]    &  1.43[9] & 700 \\

$2p_{3/2} \xrightarrow{E1}  2s_{1/2}$  & 8.40[-1]    &  2.50[9] & 500 \\
   $\ \ \ \ \ \ \ \xrightarrow{M1}  2p_{1/2}$  & 1.330     &  2.43[3]   &  \\
   $\ \ \ \ \ \ \ \xrightarrow{E2}  2p_{1/2}$  & 4.17[-3]  &  1.33[-1]   & \\
$3s_{1/2}$  $\xrightarrow{M1}  2s_{1/2}$  & 3.33[-5] &  1.22[5]    & 0.634 \\
   $\ \ \ \ \ \ \ \ \xrightarrow{E1}  2p_{1/2}$ & 2.19[-3]  &  5.14[11]  &    \\
   $\ \ \ \ \ \ \ \ \xrightarrow{E1}  2p_{3/2}$ & 4.66[-3]  &  1.06[12]  &                                \\
$3p_{1/2} \xrightarrow{E1}  2s_{1/2}$  & 1.31[-2]  &  3.74[12]  &  0.267 \\
   $\ \ \ \ \ \ \ \ \xrightarrow{M1}  2p_{1/2}$ & 1.27[-6]  & 4.15[3]        &                                  \\          
   $\ \ \ \ \ \ \ \ \xrightarrow{M1}  2p_{3/2}$ &  6.01[-5] & 1.90[5]        &                                  \\          
   $\ \ \ \ \ \ \ \ \xrightarrow{E2}  2p_{3/2}$ &  3.16[-3]  &  1.58[9]     &                                  \\ 
   $\ \ \ \ \ \ \ \ \xrightarrow{E1}  3s_{1/2}$ &   2.59[-1] &  1.90[8]   &                                    \\ 
$3p_{3/2} \xrightarrow{E1}  2s_{1/2}$  & 2.54[-2]  &  3.66[12]  &  0.273 \\

   $\ \ \ \ \ \ \ \ \xrightarrow{M1}  2p_{1/2}$ & 2.10[-5]   & 3.47[4]       &                                  \\
   $\ \ \ \ \ \ \ \ \xrightarrow{E2}  2p_{3/2}$ & 2.98[-3]   &  7.98[8]       &                                 \\
   $\ \ \ \ \ \ \ \ \xrightarrow{M1}  2p_{3/2}$ & 4.89[-5]     &  7.82[4]       &                                  \\          
   $\ \ \ \ \ \ \ \ \xrightarrow{E2}  2p_{3/2}$ & 3.10[-3]   &  7.88[8]       &                                   \\ 
   $\ \ \ \ \ \ \ \ \xrightarrow{E1}  3s_{1/2}$ & 5.21[-1]   &  3.41[8]       &                                  \\ 
   $\ \ \ \ \ \ \ \ \xrightarrow{M1}  3p_{1/2}$ &  1.332  &   6.28[1]    &                                  \\ 
   $\ \ \ \ \ \ \ \ \xrightarrow{E2}  3p_{1/2}$ &  1.54[-1]  &   1.10[-2]  &                                  \\ 
$3d_{3/2} \xrightarrow{E2}  2s_{1/2}$ & 2.48[-6]  & 4.85[3]  &  0.094 \\
   $\ \ \ \ \ \ \ \ \xrightarrow{E2}  2s_{1/2}$ & 1.16[-2]     &  4.09[9]      &                                 \\
   $\ \ \ \ \ \ \ \ \xrightarrow{E1}  2p_{1/2}$ & 7.04[-2]     &  8.87[12]      &                                 \\
   $\ \ \ \ \ \ \ \ \xrightarrow{E1}  2p_{3/2}$ & 1.43[-2]      & 1.75[12]       &                                 \\
   $\ \ \ \ \ \ \ \ \xrightarrow{E2}  3s_{1/2}$ & 9.89[-2]  &  1.67[2]        &                                   \\
   $\ \ \ \ \ \ \ \ \xrightarrow{E1}  3p_{1/2}$ & 3.38[-1]    & 2.62[7]       &                                  \\
   $\ \ \ \ \ \ \ \ \xrightarrow{E1}  3p_{3/2}$ & 6.74[-2]  &  1.39[6]      &                                  \\                     

$3d_{5/2} \xrightarrow{E2}  2s_{1/2}$  & 1.73[-2]    &  4.12[9] &  0.095 \\
   $\ \ \ \ \ \ \ \ \xrightarrow{E1}  2p_{3/2}$ & 1.29[-1]     & 1.06[13]       &                               \\
   $\ \ \ \ \ \ \ \ \xrightarrow{E2}  3s_{1/2}$ & 1.49[-1]  & 2.05[2]       &                                \\
   $\ \ \ \ \ \ \ \ \xrightarrow{E1}  3p_{3/2}$ & 6.09[-1]    & 1.33[7]       &                                 \\
   $\ \ \ \ \ \ \ \ \xrightarrow{M1}  3d_{3/2}$ & 2.392      &  2.120      &                                 \\          
   $\ \ \ \ \ \ \ \ \xrightarrow{E2}  3d_{3/2}$ & 1.82[-2]      &  2.27[-6]      &                                 \\          
$4s_{1/2} \xrightarrow{M1}  2s_{1/2}$  & 1.24[-5]      &  1.10[5]  &  1.35 \\
   $\ \ \ \ \ \ \ \ \xrightarrow{E1}  2p_{1/2}$ & 2.03[-4]      &  1.20[11]      &                                  \\
   $\ \ \ \ \ \ \ \ \xrightarrow{E1}  2p_{3/2}$ & 4.32[-4]      & 2.51[11]       &                       \\
   $\ \ \ \ \ \ \ \ \xrightarrow{E1}  3p_{1/2}$ & 1.22[-2]     & 1.20[11]       &    \\
   $\ \ \ \ \ \ \ \ \xrightarrow{E1}  3p_{3/2}$ & 2.59[-2]     &  2.48[11]      &    \\        
   $\ \ \ \ \ \ \ \ \xrightarrow{E2}  3d_{3/2}$ & 1.35[-2]   &  2.93[7]      &    \\         
   $\ \ \ \ \ \ \ \ \xrightarrow{E2}  3d_{5/2}$ & 2.05[-2]  &  4.40[7]      &    \\
\end{tabular}
\end{ruledtabular}
\label{tab4}
\end{table}

\begin{table*}
\caption{Variation in the lifetimes of different states in Li and Ca XVIII against the $D$ values obtained using Model B.}\label{tab5}
\begin{center}
\begin{ruledtabular}
\begin{tabular}{c c c c c c c c c c c c}
      &\multicolumn{5}{c}{Li (in $ns$)}                      &    &\multicolumn{5}{c}{Ca XVIII (in $ps$)}\\
\cline{2-6}\cline{8-12} & & \\
D     & 2p$_{1/2}$ & 3s$_{1/2}$ &  3p$_{1/2}$  & 3d$_{3/2}$ & 4s$_{1/2}$ & D     & 2p$_{1/2}$& 3s$_{1/2}$ & 3p$_{1/2}$ & 3d$_{3/2}$ & 4s$_{1/2}$\\      
200   & 26.60     & 30.38    & 202.19      & 14.63     & 56.36     & 7.52  &  741     & 1.09     &  0.431    & 0.145     & 1.64 \\
100   & 26.18     & 31.09    & 195.55      & 14.76     & 57.47     & 1.50  &  494     & 1.28     &  0.520    & 0.690     & 2.56  \\
50    & 25.09     & 33.23    & 190.93      & 15.34     & 62.61     & 1.00  &  334     & 1.55     &  0.661    & 0.208     & 4.83  \\
30    & 24.45     & 36.90    & 181.53      & 16.63     & 73.11     & 0.80  &  242     & 1.88     &  0.851    & 0.264     & 12.5  \\
20    & 23.31     & 44.36    & 184.68      & 19.63     & 99.33     & 0.70  &  192     & 2.22     &  1.075    & 0.339     & 82.8  \\    
15    & 22.46     & 56.39    & 191.07      & 25.34     & 139.0     & 0.60  &  140     & 2.94     &  1.620    &           & 462   \\
12    & 21.75     & 75.65    & 205.03      & 38.84     & 195.3     & 0.50  &  91      & 5.11     &  4.933    &           & 556   \\
10    & 21.05     & 109.80   & 235.49      &           &           &       &          &          &           &           &       \\
09    & 20.73     & 147.75   & 270.92      &           &           &       &          &          &           &           &      \\

\end{tabular}
\end{ruledtabular}
\end{center}
\end{table*}
\clearpage
 
\begin{figure*}
\includegraphics{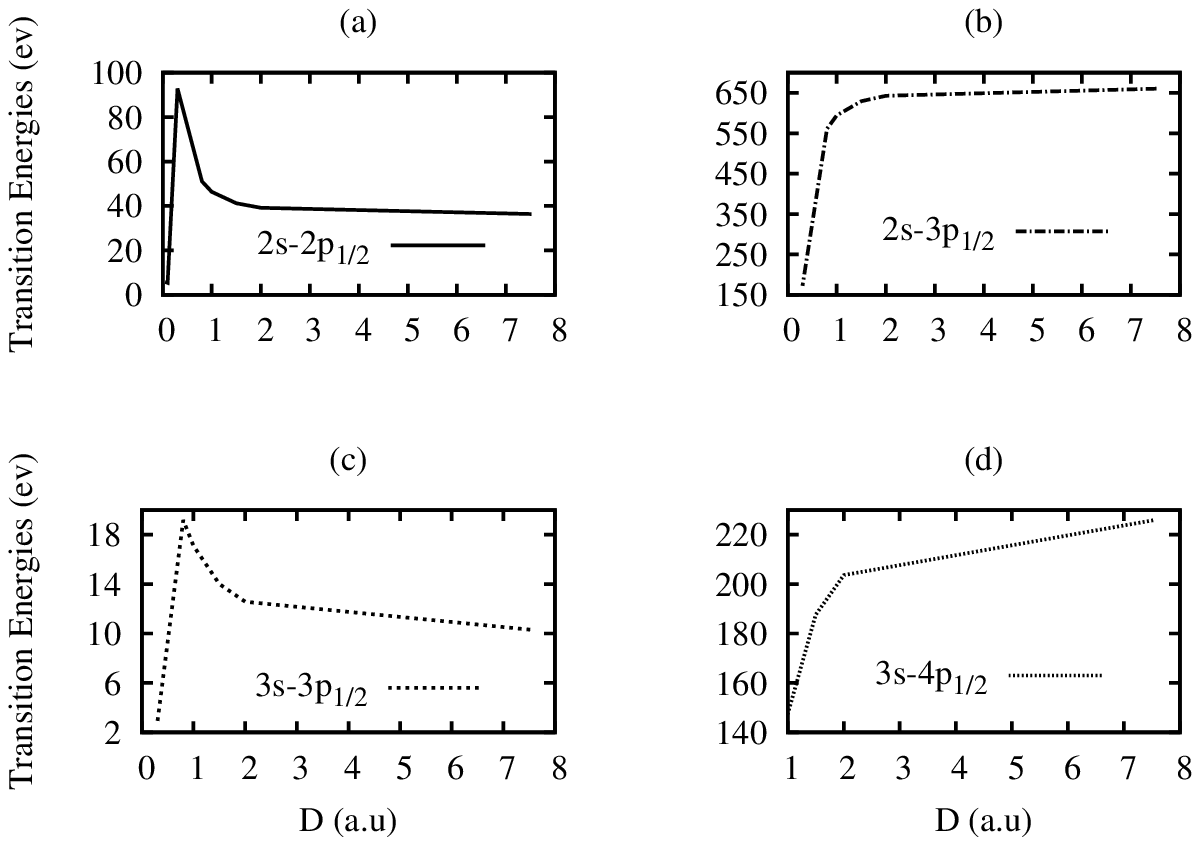}
\caption{Variation in the transition energies with Debye screening length in Ca XVIII using Model B. Panels (a) and (b) 
correspond to the 2s$_{1/2}$-2p$_{1/2}$ and 2s$_{1/2}$-3p$_{1/2}$ transitions, respectively, and panels (c) and (d) 
correspond to the 3s$_{1/2}$-3p$_{1/2}$ and 3s$_{1/2}$-4p$_{1/2}$ transitions, respectively.}
\end{figure*}

\begin{figure*}
  \includegraphics[width=18.4cm, height=8.7cm]{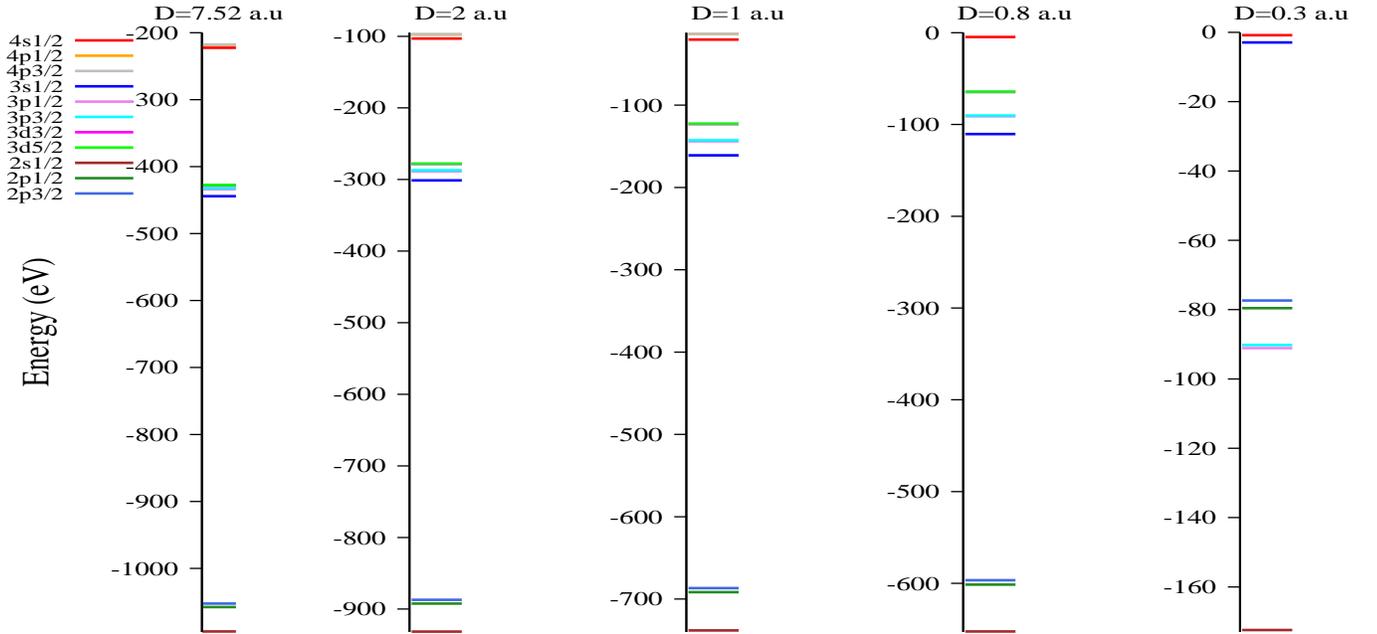}
\caption{(Color online) Grotrian diagram of energy levels in Ca XVIII for different values of $D$ from Model B.}
\label{fig8}
\end{figure*}

\begin{figure*}
\includegraphics{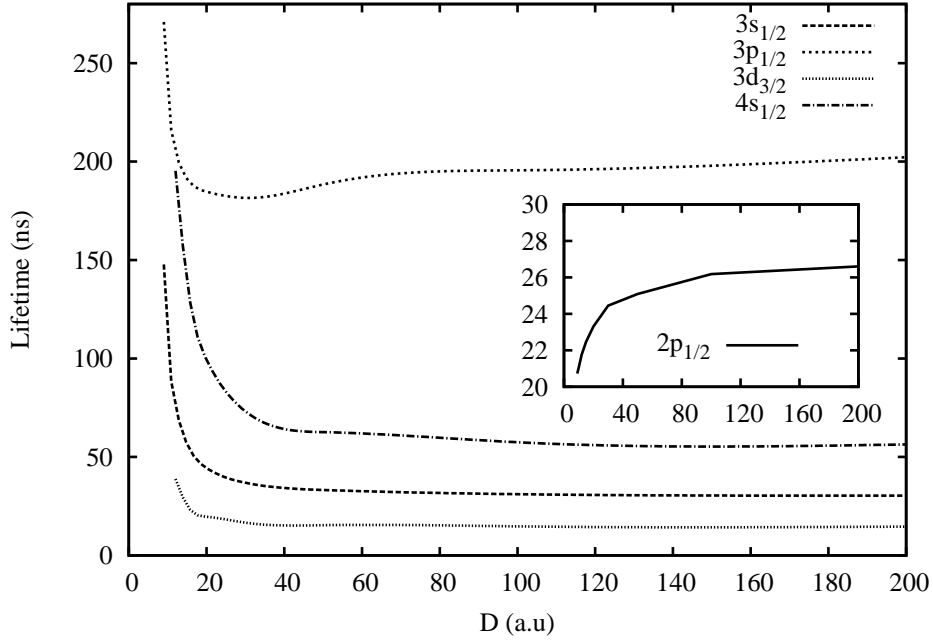}
\caption{Variation in the lifetimes of the excited 3s$_{1/2}$, 3p$_{1/2}$, 3d$_{3/2}$ and 4s$_{1/2}$ states in Li I 
with $D$ values with Model B. The inset plot is for the 2p$_{1/2}$ state.}
\label{fig9}
\end{figure*}

\begin{figure*}
\includegraphics{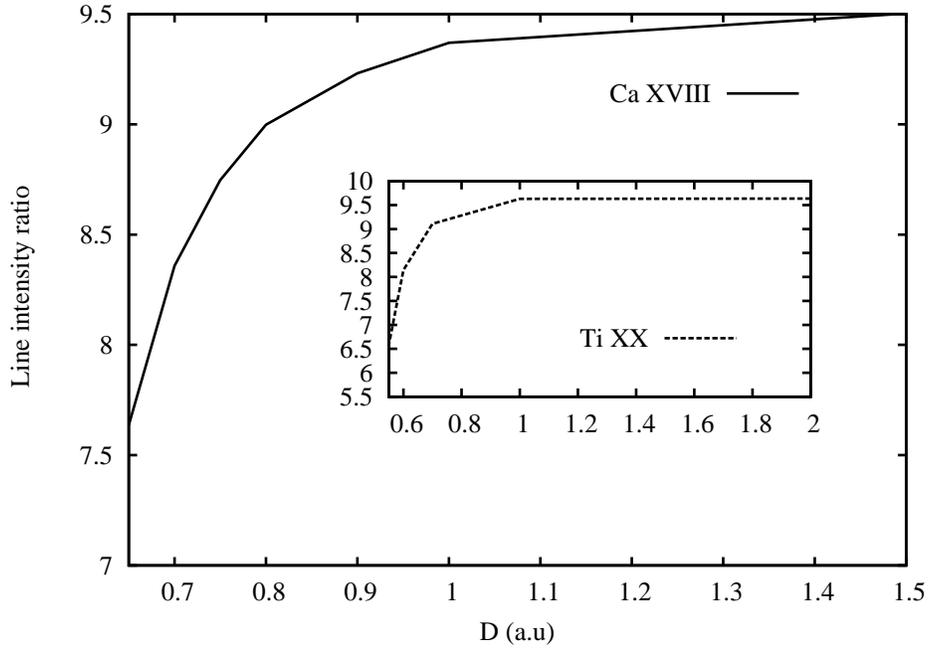}
\caption{Variation in the line intensity ratio of the $3d ^{2}D_{3/2} \rightarrow 2p ^{2}P_{1/2}$ and 
$3s ^{2}S_{1/2} \rightarrow 2p ^{2}P_{3/2}$ transitions of Ca XVIII with $D$. The same is shown in the inset plot for Ti XX.}
\label{fig10}
\end{figure*}

\end{document}